\documentclass[12pt]{article}
\usepackage{amsmath,epsfig}
\usepackage{amsmath}
\usepackage{amssymb}
\usepackage{mathrsfs}
\usepackage{amsmath,epsfig}
\usepackage{graphicx}
\usepackage{amsmath,epsfig}
\usepackage{epsfig}
\newsavebox{\DSLASH}
\sbox{\DSLASH}{$D$\hspace{-2.5mm}/}

\begin{document}

\vspace{4cm}
\begin{center}
{\large \bf{ Heavy Charged Gauge Bosons with General CP Violating Couplings }}\\
\vspace{2cm}

Mojtaba Mohammadi Najafabadi \footnote{Email:
mojtaba@mail.ipm.ir}\\
{\sl School of Particles and Accelerators, \\
Institute for Research in Fundamental Sciences (IPM)\\
P.O. Box 19395-5531, Tehran, Iran}\\

\vspace{3cm}
 \textbf{Abstract}\\
 \end{center}

Heavy gauge bosons such as $W^{\prime}$ are expected to
exist in many extensions of the Standard Model. In this paper,
it is shown that the most general Lagrangian
for the interaction of $W^{\prime}$ with top and bottom quarks
which consists of V-A and V+A structure with in general complex
couplings produces an Electric Dipole Moment (EDM) for the top quark at
one loop level.
We predict the allowed ranges for the mass and couplings of $W^{\prime}$
by using the upper limit on the top quark EDM.

\newpage
\section{Introduction}

The Standard Model (SM) of the particles has been found to be in
a good agreement with the present experimental data in many of
its aspects. Nonetheless, it is believed to leave many questions
unanswered, and this belief has resulted in numerous theoretical
and experimental attempts to discover a more fundamental
underlying theory. Various types of experiments may expose the
existence of physics beyond the SM, including the search for
direct production of exotic particles at high energy colliders. A
complementary approach in hunting for new physics is to examine
its indirect effects in higher order processes.

There are many different models which predicts the existence of
new charged gauge bosons, $W^{\prime}$. These scenarios include
the Little Higgs model \cite{lh1} \cite{lh2}, Grand Unified
Theories \cite{gut}, Universal Extra Dimension \cite{ued},
Left-Right Symmetric Model \cite{lrsm1} and some other models.
One must note that the properties and interactions of $W^{\prime}$
depend on the model. One of the simple extension of the SM is the
Left-Right Symmetric Model. It is based on the $SU(2)_{R} \times
SU(2)_{L}\times U(1)$ gauge group. The new $SU(2)_{R}$ symmetry
leads to additional $W^{\prime},Z^{\prime}$ gauge bosons. For a
detailed discussion of Left-Right Symmetric Model see for example
\cite{lrsm1},\cite{lrsm2},\cite{lrsm3}. The Left-Right Symmetric
Model is constructed by placing the fermion right-handed singlets
into doublets regarding $SU(2)_{R}$. This requires to introduce
right-handed neutrinos. One the interesting aspects of this model
is that the parity is broken spontaneously which causes to
different masses for the $SU(2)_{R}$ and $SU(2)_{L}$ gauge bosons.

Although such charged massive bosons have not been found yet
experimentally but it is widely believed that the experiments at
the LHC are able to probe them in the coming years
\cite{ptdr}, \cite{rizzo}, \cite{wp2}. 
At the LHC, for an integrated luminosity of 10 fb$^{-1}$,
$W^{\prime}$ bosons can be discovered or excluded up to a mass of
5 TeV/c$^{2}$, from an analysis of the muonic decay mode. This
result belongs to the model which makes the assumptions that the
new gauge boson $W^{\prime}$ has the same couplings as the
Standard Model $W$ boson. The capability of LHC to explore the
helicity of $W^{\prime}$ is discussed in \cite{rizzo}. There are
already direct and indirect searches for the new gauge bosons.
There is a severe limit obtained from $K_{0}-\bar{K}_{0}$ mixing:
$M_{W^{\prime}} \geq$2.5 TeV/c$^{2}$ \cite{zhang}. The direct
searches for $W^{\prime}$ can be found for example in
\cite{pdg},\cite{ds1},\cite{ds2}.

In the framework of the SM top quark is the only quark which has
a mass in the same order as the electroweak symmetry breaking
scale, $v\sim 246$ GeV, whereas all other observed fermions have
masses which are a tiny fraction of this scale. This huge mass
might be a hint that top quark plays an essential role in search
for new physics originating from physics at higher scale
\cite{beneke}. Hence, the study of interaction of top quark with
$W^{\prime}$ might give useful information about $W^{\prime}$.
For example, the interference between $W^{\prime}$ and $W$ in the
production of single top quarks is important and could be useful
in search for $W^{\prime}$ which has been discussed in \cite{wp2}.

The aim of this article is to constrain the mass of $W^{\prime}$
by considering its contribution to the electric dipole moment
(EDM) of the top quark. In \cite{toscano2}, the authors have estimated
an upper limit of $10^{-20}$ e.cm. on the top quark EDM from
the experimental bound on the neutron EDM. Combination this limit with
the contribution of the $W^{\prime}$ to top EDM leads to valuable information
on $M_{W^{\prime}}$ and its couplings.

\section{The Contribution of the $W^{\prime}$ to the Top Quark EDM}

Similar to the interaction of $Wtb$, the most general lowest
order effective Lagrangian for the interaction of $W^{\prime}$
with top and bottom quarks in the SM can be written in the
following form \cite{pdg},\cite{ds1}:
\begin{eqnarray}\label{lag}
{\cal L} = \frac{g}{\sqrt{2}}\bar{t}\gamma^{\mu}
\left( a_{L}P_{L} + a_{R}P_{R}\right)b W^{\prime}_{\mu}
\end{eqnarray}
where $P_{L}(P_{R})$ are the left-handed (right-handed) projection operators. The
$a_{L},a_{R}$ coefficients are complex in general. This signifies the CP violating
effects. In this notation, $a_{L} = 1$ and $a_{R} = 0$ for a so-called
SM-like $W^{\prime}$.

It is worth mentioning that in Eq.\ref{lag} if we replace
$W^{\prime}$ gauge boson by the Standard Model $W$ gauge boson,
from the B decay processes the limits on $a_{L},a_{R}$ are:
$Re(a_{R})\leq 4\times 10^{-3}$, $Im(a_{R})\leq 10^{-3}$ and
$Im(a_{L})\leq 3 \times 10^{-2}$ \cite{b1},\cite{b2},\cite{b3}.

\begin{figure}
\centering
  \includegraphics[width=10cm,height=6cm]{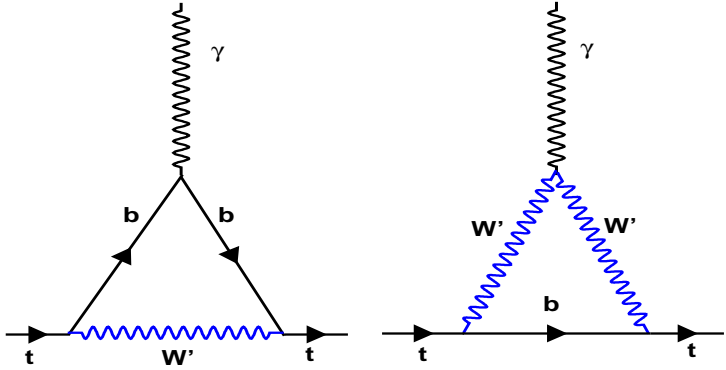}\\
  \caption{Feynman diagrams contributing to the on shell $t\bar{t}\gamma$.}\label{vertex}
\end{figure}

The introduced Lagrangian in Eq.\ref{lag} induces an electric
dipole moment for the top quark at the one loop level via the
Feynman diagrams shown in Fig.\ref{vertex}. One should note that
all the particles are taken on-shell. After calculation of the
one loop corrections to the vertex of $\bar{t}t\gamma$ shown in
Fig.\ref{vertex}, we find some terms with different structures.
The coefficient of the structure of
$\sigma_{\mu\nu}\gamma_{5}q^{\nu}$ gives the top quark electric
dipole moment where $q^{\nu}$ is the four momentum of photon
\cite{edm0},\cite{edm1}. It should be noted that this structure arises via
radiative corrections and does not exist at tree level.

After all calculation, the top EDM is found as:
\begin{equation}
\label{topEDM} d_{t}=-\frac{e}{m_{W^{\prime}}}\frac{3\,\alpha}{32
\pi}\,\frac{m_{b}}{m_{W^{\prime}}}\,\left(V_{1}(x_{b},x_{W^{\prime}})+
\frac{1}{3}\,V_{2}(x_{b},x_{W^{\prime}})\right)\,{\text Im}\left(a_{L} a^{*}_{R}\right),
\end{equation}
\noindent where $x_{a}=m_a^{2}/m_t^{2}$. The
$V_{1,2}$ are the functions stand for the contribution of the Feynman
diagram where the photon emerges from the $W^{\prime}$ boson and the $b$
quark line, respectively. They have the following forms:
\begin{eqnarray}
V_{1}=-\left(4x_{W^{\prime}}-x_{b}+1\right)f(x_{b},x_{W^{\prime}})-
\left(x_{b}^2+4x_{W^{\prime}}^2-5x_{b}x_{W^{\prime}}-3x_{W^{\prime}}-2x_{b}+1\right)g(x_{b},x_{W^{\prime}})\nonumber \\
V_{2}=-\left(4x_{W^{\prime}}-x_{b}+1\right)f(x_{W^{\prime}},x_{b})+
\left(x^{2}_{b}+4x_{W^{\prime}}^2-5x_{b}x_{W^{\prime}}-3x_{W^{\prime}}-2x_{b}+1\right)g(x_{W^{\prime}},x_{b})
\end{eqnarray}
\noindent where the functions of $f$ and $g$ are as follows:
\begin{eqnarray}
f(a,b)&=&\left(\frac{1+a-b}{2}\right)\log\left(\frac{b}{a}\right)+\sqrt{(1-a-b)^2-4ab}\,\times{\rm
ArcSech}\left(\frac{2 \sqrt{ab}}{a+b-1}\right)+2 \nonumber \\
g(a,b)&=&-\frac{1}{2}\log\left(\frac{b}{a}\right)-\frac{1+a-b}{\sqrt{(1-a-b)^2-4ab}}\,\times{\rm
ArcSech}\left(\frac{2 \sqrt{ab}}{a+b-1} \right) \nonumber
\end{eqnarray}

\section{Results}

\begin{figure}
\centering
  \includegraphics[width=11cm,height=8cm]{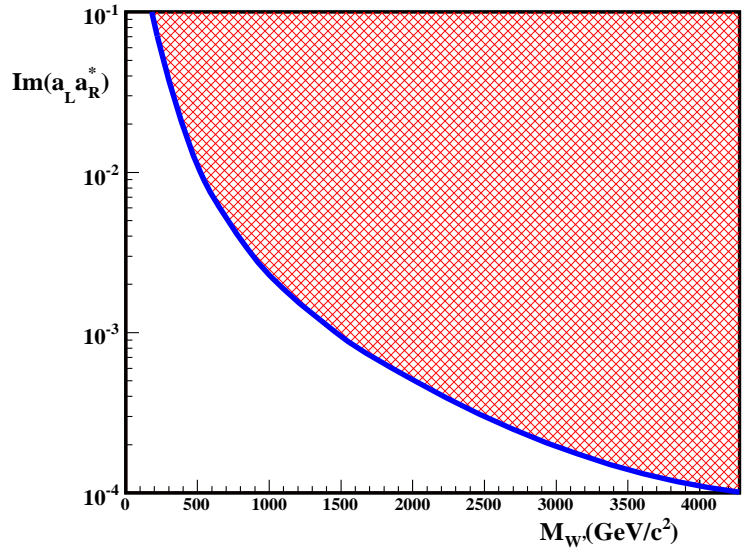}\\
  \caption{The shaded region is the excluded region for
the $W^{\prime}$ mass and $Im(a_{L} a^{*}_{R})$ by this analysis.}\label{exclusion}
\end{figure}

In \cite{toscano2}, the authors have predicted an
upper bound for the top quark EDM using
the experimental limit on the neutron EDM.
Their estimate for the top quark EDM is $10^{-20}$ e.cm.
In Eq.\ref{topEDM}, if we assume $Im\left(a_{L} a^{*}_{R}\right) \sim 10^{-1}$
and by using the bound of the top EDM, the upper limit of $190$ GeV/c$^{2}$ is achieved
for the mass of $W^{\prime}$ and if $Im\left(a_{L} a^{*}_{R}\right) \sim 10^{-3}$
we have $M_{W^{\prime}} \leq 1470$ GeV/c$^{2}$. The shaded region in Fig.\ref{exclusion}
is the excluded region in the plane of $M_{W^{\prime}}$ and  $Im\left(a_{L} a^{*}_{R}\right)$.
Fig.\ref{exclusion} obviously presents the the strong dependence of the upper bound of the
mass of $W^{\prime}$ on the $Im\left(a_{L} a^{*}_{R}\right)$.

The predicted lower limit for the $W^{\prime}$ mass from other
studies ($K_{0}-\bar{K}_{0}$ mixing) which mentioned in the
introduction can be used to estimate the allowed range for
$Im\left(a_{L} a^{*}_{R}\right)$. In Eq.\ref{topEDM} if we put
$d_{t} < 10^{-20}$ and $M_{W^{\prime}} \geq$ 2.5 TeV/c$^{2}$ the
upper bound of $3.18 \times 10^{-4}$ is derived for $Im\left(a_{L}
a^{*}_{R}\right)$.

In \cite{ds1} a search has been performed for $W^{\prime}$ bosons
which decay to $t+b$, using 0.9 fb$^{-1}$ of data recorded by D0
detector in proton anti-proton collisions. A 95$\%$ C.L. upper
limit on $\sigma(p\bar{p}\rightarrow W^{\prime})\times
BR(W^{\prime}\rightarrow tb)$ has been set. This excludes the
gauge couplings ($a_{L},a_{R}$) above $\sim 0.7$ for $W^{\prime}$
bosons with a mass of 600 GeV/c$^{2}$. From the current analysis,
for the $W^{\prime}$ bosons with a mass of 600 GeV/c$^{2}$,
$Im\left(a_{L} a^{*}_{R}\right)$ above $\sim 0.007$ is excluded.

\section{Conclusion}
In this paper we focus our attention on the contribution of the
$W^{\prime}$ gauge boson to the electric dipole moment (EDM) of
the top quark. The most general Lagrangian for the interaction of
$W^{\prime}$ with top and bottom quarks which consists of V-A and
V+A structure with in general complex couplings $(a_{L},a_{R})$
produces an EDM for the top quark at level of one loop. The top
EDM is proportional to $Im\left(a_{L} a^{*}_{R}\right)$. Using
the upper limit on the top EDM, we exclude the region shown in
Fig.\ref{exclusion} in the plane of $M_{W^{\prime}}$ and
$Im\left(a_{L} a^{*}_{R}\right)$. For example, for $Im\left(a_{L}
a^{*}_{R}\right) \sim$ 0.001, the $W^{\prime}$ boson mass above
1470 GeV/c$^{2}$ is excluded. The upper bound of $3.18 \times
10^{-4}$ is derived for $Im\left(a_{L} a^{*}_{R}\right)$ by
considering the lower limit on the $M_{W^{\prime}}$ from
$K_{0}-\bar{K}_{0}$ mixing studies.

{\large \bf Acknowledgments}\\
The author would like to thank B. Safarzadeh for reading the manuscript.\\

\end{document}